\documentclass[12pt, twoside]{article}
\usepackage{a4wide,amssymb,cite,amsmath,setspace,graphicx,subfigure}
\usepackage{epsfig}

\newcommand{\be}{\begin{equation}}
\newcommand{\ee}{\end{equation}}
\newcommand{\bea}{\begin{eqnarray}}
\newcommand{\eea}{\end{eqnarray}}

\newcommand{\dis}[1]{\begin{equation}\begin{split}#1\end{split}\end{equation}}
\newcommand{\eq}[1]{Eq.~(\ref{#1})}
\newcommand{\bfrac}[2]{{\left(\frac{#1}{#2} \right)  }}
\newcommand{\VEV}[1]{{\langle{#1}\rangle  }}

\newcommand{\stau}{{\tilde{\tau}_1 }}
\newcommand{\gravitino}{{\tilde{G}}}
\newcommand{\mstau}{{m_{\stau}}}
\newcommand{\mgravitino}{{m_{3/2}}}

\newcommand{\tgbeta}{{\tan\beta}}

\newcommand{\Mp}{M_P}
\newcommand{\tev}{\,\textrm{TeV}}
\newcommand{\gev}{\,\textrm{GeV}}

\begin{document}

\begin{titlepage}

\rightline{PNUTP-10/A02}
\rightline{KIAS-P1005}
\rightline{CERN-PH-TH/2010-038}

\begin{centering}
\vspace{1cm}
{\large {\bf  Dark matter, $\mu$ problem and neutrino mass with gauged R-symmetry}}\\

\vspace{1.5cm}

 {\bf Ki-Young Choi} $^{a,*}$, {\bf Eung Jin Chun} $^{b,**}$, {\bf Hyun Min Lee} $^{c,***}$
\\
\vspace{.2in}

$^{a}$ {\it Department of Physics, Pusan National University,
Busan 609-735, Korea.}

\vspace{3mm}

$^{b}$ {\it Korea Institute for Advanced Study, Hoegiro 87, Dongdaemoon-gu, Seoul 130-722, Korea.}

\vspace{3mm}

$^{c}$ {\it CERN, Theory division, CH-1211 Geneva 23, Switzerland.}

\vspace{.1in}

\end{centering}

\begin{abstract}
\noindent
We show that the $\mu$ problem and the strong CP
problem can be resolved in the context of the gauged $U(1)_R$
symmetry, realizing an automatic Peccei-Quinn symmetry. In this
scheme, right-handed neutrinos can be introduced to explain small
Majorana or Dirac neutrino mass. The $U(1)_R$ D-term mediated SUSY
breaking, called the $U(1)_R$ mediation, gives rise to a specific
form of the flavor-conserving superpartner masses. For the given
solution to the $\mu$ problem, electroweak symmetry breaking
condition requires the superpartners of the Standard Model at low
energy to be much heavier than the gravitino. Thus dark matter
candidate can be either gravitino or right-handed sneutrino. In
the Majorana neutrino case, only gravitino is a natural dark
matter candidate. On the other hand, in the Dirac neutrino case,
the right-handed sneutrino can be also a dark matter candidate as
it gets mass only from SUSY breaking. We discuss the non-thermal
production of our dark matter candidates from the late decay of
stau and find that the constraints from the Big Bang
Nucleosynthesis can be evaded for a TeV-scale stau mass.

\end{abstract}

\vspace{1cm}

\begin{flushleft}

$^{*}~$  e-mail address: kiyoung.choi@pusan.ac.kr\\
$^{**}$  e-mail address: ejchun@kias.re.kr  \\
$^{***}$  e-mail address: hyun.min.lee@cern.ch \\

\end{flushleft}

\end{titlepage}

\section{Introduction}

$N=1$ supersymmetry (SUSY) contains a continuous $U(1)_R$ group
transforming different supercharges and thus distinguishing
between bosonic and fermionic components of superfields. A
discrete subgroup of it, called R-parity, can remain respected
after SUSY breaking, and commonly used to explain the stability of
proton. In local supersymmery (supergravity), such an R-symmetry,
either discrete or continuous, is gauged and should respect
anomaly-free condition \cite{Ibanez91, Chamseddine95}. A gauged
R-symmetry controls all the fields in a theory and thus its
anomaly-free condition is very restrictive.  Due to this property,
R-symmetries can provide a powerful tool for phenomenological
applications such as $U(1)_R$ as a family symmetry \cite{Chun95},
as a resolution to the $\mu$ problem and $B$/$L$ conservation
\cite{Nilles96}, as an origin of supersymmetry breaking
\cite{Kitazawa00} and the D-term inflation \cite{DTinf}. More
recently, an $U(1)_R$ mediated supersymmetry breaking model has
been constructed based on a six-dimensional flux compactification
\cite{6Dfluxcompact} and its phenomenological application was
investigated by some of the authors \cite{U1R,U1Rpheno}. In 6D
compactifications, even if the hidden sector is geometrically
separated from the visible sector, the hidden sector SUSY breaking
is generically not sequestered \cite{faleelud}, as compared to the
5D counterpart. However, it has been shown that the moduli F-term
contribution to the soft mass is cancelled by the hidden F-term
contribution such that the $U(1)_R$ mediation can be dominant
\cite{U1R,U1Rpheno}.

In this paper, extending the previous studies \cite{U1Rpheno}, we
propose a resolution of the $\mu$ problem along the line of
Ref.~\cite{Kim84}, realizing also the axion solution to the strong
CP problem \cite{Kim87}. At the same time,  the observed neutrino
masses and mixing can be explained by introducing three
right-handed neutrinos, which can form Majorana neutrinos by the
usual seesaw mechanism (at the intermediate axion scale of order $
10^{10-12}$ GeV or at the TeV scale) or Dirac neutrinos with tiny
neutrino Yukawa couplings. In this framework, we find that the
electro-weak symmetry breaking (EWSB) condition requires a
peculiar superparticle mass spectrum: all the superparticles in
the minimal supersymmetric standard model (MSSM) sector have
masses of the TeV scale whereas the gravitino or right-handed
sneutrino masses can be around 100 GeV. This is in contrast to the
previous study of the $U(1)_R$ phenomenology \cite{U1Rpheno} where
the gaugino masses are assumed to be comparable  to the scalar
soft masses at the GUT scale and the $\mu$ and $B\mu$ terms are
assumed to be given such that the EWSB condition is satisfied.

As the superpartners in the MSSM turn out to be heavy, either the
gravitino or the right-handed (RH) sneutrino can be a natural dark
matter candidate. We focus on the non-thermal productions of dark
matter depending on the nature of neutrinos and impose the Big
Bang Nucleosynthesis (BBN) constraints. When neutrinos are of
Majorana type, it is typical that gravitino is the lightest
supersymmetric particle (LSP) and stau is the next LSP (NLSP). In
this case, the stau decay after the freezeout becomes a dominant
source for the non-thermal production of gravitino relic density.
On the other hand, when neutrino is of Dirac type, RH sneutrino
can be the LSP and then gravitino is the NLSP. Because of the
tiny Dirac neutrino Yukawa couplings, the thermal production of the RH
sneutrino from the decay of heavy superparticles is usually
suppressed. Then, the decay of stau can be a dominant source for
the RH sneutrino relic density. We point out that in both
gravitino and RH sneutrino dark matter scenarios, the BBN problem
coming from the late decay of stau can be avoided for the
TeV-scale stau mass as required from the EWSB conditions.

The paper is organized as follows. First, we present a 4D
effective theory with gauged $R$-symmetry where the hidden sector
SUSY breaking is introduced and the visible sector contains no
additional representations under the standard model gauge group
other than the MSSM content. In the next section, we show that the
soft mass parameters at the GUT scale are determined from the
$U(1)_R$ mediation and the $\mu$ term and the neutrino masses are
dynamically generated with nonzero singlet vacuum expectation
values (VEVs). Consequently, deriving the low-energy SUSY spectrum, we
discuss  the relic density of dark matter and the BBN constraints,
depending on the nature of neutrinos. Finally, conclusion is
drawn. There are four appendices comprised of the minimization of
the singlet scalar potential, the identification of the axion for
multiple scalar VEVs, the discussions on the effective $R$-parity
and PQ symmetry violating terms induced after the $R$-symmetry
breakdown.

\section{A SUSY model with gauged R-symmetry}

We first review the consistency of the Fayet-Iliopoulos(FI) term in supergravity 
and explain the 4D effective gauge supergravity
recently derived from a six-dimensional flux compactification. For
generation-independent $R$-charges and renormalizable Yukawa
couplings in the MSSM, the gauged $U(1)_R$ is anomaly-free up to a
Green-Schwarz counter term. We obtain the $\mu$ term and the
Majorana/Dirac neutrino mass terms through the superpotential
terms with singlets of intermediate-scale VEVs. In both Majorana
and Dirac neutrino cases, we present the representative models
where negative $R$-charges  are assigned for the MSSM scalar
partners except for the Higgs doublets.

\subsection{Fayet-Iliopoulos term in 4D supergravity}

It has been recently pointed out that a $U(1)$ gauge theory with constant FI term can be consistent with supergravity, provided that an exact global symmetry is present \cite{FIsugra}. However, any global symmetry is believed to be broken in quantum gravity,
leading to a conclusion that there is no consistent $U(1)$ with a constant FI term.
Here we review the FI term in supergravity focusing on the chiral compensator formalism discussed by Komargodski and Seiberg in Ref.~\cite{FIsugra} and comment on the consistency of the FI term for the gauged $U(1)_R$.

The general 4D gauged supergravity action in Weyl compensator
formalism is 
\be S=\int d^4 x \Big[d^4\theta {\bf E}(-3C^\dagger
e^{2\xi g_RV_R/3} C \,e^{-K_0(\Phi^\dagger_i/C, \Phi_i/C)/3})+\int d^2\theta {\cal E} C^3 W(\Phi_i/C) +{\rm
h.c.}\Big] \ee 
where $C$ is the compensator superfield, which
becomes $C=C_0+\theta^2 F_C$ in super-Weyl gauge, ${\bf E}$ and
${\bf {\cal E}}$ are the full and chiral superpace measures,
respectively, $V_R$ is the $U(1)_R$ vector superfield, and  $g_R$ is the $U(1)_R$ gauge coupling.
For the gravitino of $R$-charge $+1$,  the constant Fayet-Iliopoulos(FI) term for the gauged $U(1)_R$ is quantized as $\xi=2$.
The above supergravity action can be made $U(1)_R$ gauge invariant by a super-Weyl transformation \cite{DTinf,U1R}. 
A construction of the gauged $U(1)_R$ invariant action in 4D
supergravity has been originally done in
Ref.~\cite{u1rsupergravity}.

The $U(1)_R$ transformation, which should not be confused with the gauged R-symmetry, is defined such that all the chiral superfields have R-charge 
$2/3$. We note that the fermionic superpartners have R-charge $-1/3$. 
The original theory with general superpotential  $W(\Phi_i)$ is made R-symmetric by adding an additional chiral superfield $C$ with R-charge $2/3$.

Now we consider a local $U(1)_Q$ without FI term under which the $U(1)$ charges are $Q[\Phi_i]=Q_i$ and $Q[C]=0$. After introducing a nonzero FI term for the $U(1)_Q$ in the K\"ahler potential,
we need to make charge shifts, obtaining a new local $U(1)_{\bar Q}$: $U(1)$ charges are ${\bar Q}[C]=-\xi/3$ and ${\bar Q}[\Phi_i]=Q_i+R_i\xi/2 -\xi/3$ where
$R_i$ are new R-charges satisfying $\sum_i R_i=2$ for chiral superfields appearing in each term of the superpotential.

After $C$ gets a nonzero vacuum expectation value, $C=C^\dagger=M_P$,  the $U(1)_{\bar Q}$ and the $U(1)_R$ is broken down to a gauged R-symmetry,  $U(1)_{\bar Q}+\frac{\xi}{2}U(1)_R\equiv U(1)_{\bar R}$. The R-charges of this gauged R-symmetry are ${\bar R}[C]=0$,
${\bar R}[\Phi_i]=Q_i+ R_i \xi/2\equiv {\bar R}_i$. The R-charges of the fermionic superpartners of $\Phi_i$ are ${\bar R}[\psi_i]=Q_i+(R_i-1)\xi/2$.
Therefore, since $\sum_i {\bar R}_i=\xi$ and $\sum_i R_i=2$ for chiral superfields appearing in each term of the superpotential, one can draw a conclusion that $\sum_i Q_i=\sum_i {\bar R}_i-(\xi/2)\sum_i R_i=0$, so there appears a global symmetry for the general superpotential. 

This result is due to the assumption that there is a local $U(1)_Q$ in the limit of a vanishing FI term. In order to construct the $U(1)$ theory with nonzero FI term, however, one only has to start with $U(1)_{\bar Q}$ symmetry instead of local $U(1)_Q$. Furthermore, when the FI term is quantized as required for charge quantization, there is no limit of a vanishing FI term. It has been shown that a consistent 4D supersymmetric vacuum with gauged R-symmetry can be obtained below the compactification scale in six-dimensional gauged supergravity \cite{6Dfluxcompact,U1R}. In this case, the quantization of the FI term is originated from the flux quantization in extra dimensions.
For instance, when $\xi=2$, we only have to start with the $U(1)_{\bar Q}$ having charges
${\bar Q}_i={\bar R}_i-\frac{2}{3}$ in terms of the R-charges of the gauged R-symmetry, while the $U(1)_Q$ symmetry does not need to be imposed. 
The charges of the would-be global symmetry $U(1)_Q$ are $Q_i={\bar R}_i-R_i$, so $U(1)_Q$ would be unbroken only if one can find $R_i$'s satisfying $\sum_i R_i=2$ at all orders.
However, it is also possible to have $\sum_i {\bar R}_i=2$ but $\sum_i R_i\neq 2$ for higher dimensional terms in the superpotential.

As will be discussed in later sections, the PQ symmetry appears in our model and it is nothing but a global symmetry with new R-charges given by ${\tilde r}_i=r_i+q_i$ with $q_i$ being the PQ charges and $r_i$ being the R-charges of the gauged $U(1)_R$. Here, $\sum_i {\tilde r}_i=2$ is guaranteed at low orders in the superpotential by the fact that $\sum_i q_i=0$ and $\sum_i r_i=2$.
However, at higher orders, even for $\sum_i r_i=2$, we found that $\sum_i q_i =0$ or $\sum_i {\tilde r}_i=2$ is not satisfied any more.  For instance, for $X,Y$ singlets considered in our previous paper, the PQ symmetry is an accidental symmetry which holds for the quark/lepton Yukawa couplings and the $\mu$ term.
However,  the $PQ$ symmetry is broken by the other Planck-scale suppressed $U(1)_R$-invariant interactions.

\subsection{4D effective supergravity from a 6D flux compactification}

In this section, we consider a concrete form of the K\"ahler potential and the
superpotential  derived from a flux compactification in six
dimensions \cite{U1R,U1Rpheno}. The bulk theory is based on a 6D
chiral gauged supergravity constructed by Nishino and Sezgin
\cite{6dgsugra}. The gauged $U(1)_R$ appears as a partial gauging
of the bulk $R$-symmetry. In this flux compactification, there is
a non-vanishing gauge flux along the $U(1)_R$, making the 4D
Minkowski space flat while the 2D extra dimensions are
compactified on the sphere with a wedge cut out
\cite{sscompact,6Dfluxcompact}. There are two 3-branes with
nonzero equal tension at the poles of the wedged sphere so visible
sector fields are located at one pole  and the hidden sector
fields are located at the other pole. In order to stabilize the
remaining modulus, some bulk dynamics should be taken into account
too.

In the 4D effective supergravity, the FI term is given by $\xi=2$ and part of the K\"ahler potential without FI term is
\bea
K_0&=&-\ln\Big(\frac{1}{2}(S+S^\dagger)\Big) \nonumber \\
&&-\ln\Big(\frac{1}{2}(T+T^\dagger-8g_R V_R)-Q^\dagger_i e^{-2r_i g_R V_R}Q_i-Q^{'\dagger}e^{-4g_R V_R}Q'
-\varphi^\dagger e^{-2r_\varphi g_R V_R}\varphi\Big) \nonumber \\
&&+M^\dagger e^{-2r_M g_R V_R}M.
\eea
Here $S,T$ are the moduli that mix the dilaton and the volume modulus,
$Q_i$ are visible brane fields, $Q',\varphi(M)$ are hidden sector fields living on the hidden brane (in bulk). In our model, the gauged R-symmetry is spontaneously broken by a flux compactification in six dimensions  \cite{U1R,U1Rpheno}. So it is non-linearly realized by the axion of the bulk T-modulus and the mass of the $U(1)_R$ gauge boson is of order the compactification scale.
This is manifest in the above form of the K\"ahler potential.

The superpotential for the modulus and the hidden sector is \be
W_{\rm moduli}= W_0+fQ'
+\frac{\lambda}{M^n}e^{-bS}+\lambda'\varphi^p M^2+\kappa\varphi^q
\label{modulispot} \ee where the $R$-charges are $r_{Q'}=2$,
$r_M=-\frac{2}{n}$ and $r_\varphi=\frac{2}{q}=\frac{2(n+2)}{pn}$.
Here we introduced the uplifting sector parametrized by $f$ and the
bulk sector responsible for the gaugino condensate, in an
explicitly $U(1)_R$ invariant fashion. 
When a Green-Schwarz coupling to the $T$-modulus is responsible for cancelling the $R$-symmetry anomalies, the superpotential for the gaugino condensate would get a $T$-dependent factor with $S$ being replaced by $S+\epsilon T$. 
But we assume $\epsilon\ll 1$ such that the $T$-modulus dependence gives a negligible effect on the soft scalar masses.

On the other hand, $W_0$ stands for a nonzero VEV of the superpotential obtained
after a spontaneous breaking of the $U(1)_R$ in the hidden sector.
In generalized O'Raifeartaigh model with renormalizable
interactions, independent of the $R$-symmetry breaking, the
superpotential VEV is undetermined at tree level as it is
proportional to the pseudo-moduli \cite{generalOR}. However, it is
possible to stabilize the pseudo-moduli at a nonzero value from
the Coleman-Weinberg potential at one-loop \cite{shih}. In this
case, the superpotential can get a nonzero VEV, $W_0\neq 0$. 
However, the $R$-symmetry breaking sector giving rise to $W_0\neq 0$ would necessarily break SUSY and generate a positive vacuum energy, because of the consistency condition, $2W_0=\sum_i r_i\phi_i \frac{\partial W}{\partial\phi_i}$.
For instance, a bulk $R$-symmetry breaking field $\Phi$ with R-charge $+2$ leads to the scalar potential, 
$V_{\Phi}=\frac{1}{({\rm Re}S) ({\rm Re} T-Q^\dagger Q)}|F_\Phi|^2$.
Because of the bound \cite{dine} on $|W_0|$: $2|W_0|\leq f_r F$ with $f_r\equiv\sum_i r^2_i |\phi_i|^2$ and $F^2\equiv\sum_i|\frac{\partial W}{\partial\phi_i}|^2$, the positive vacuum energy could not be cancelled by the perturbative contribution proportional to $|W_0|^2$ for $f_r\ll M_P$. However, since the R-symmetry anomalies are cancelled by a Green-Schwarz mechanism, a non-perturbative correction may break the R-symmetry dynamically so that we may avoid the no-go theorem based on the perturbative generation of the superpotential. 
In this case, the order parameter of the R-symmetry breaking is now the superpotential vev itself, not the vev of a fundamental scalar field.

The moduli stabilization with the effective superpotential
(\ref{modulispot}) has been discussed in Ref.~\cite{U1Rpheno}. The
real part of the $T$ modulus was shown to be stabilized at
$t\simeq 1$ mainly by the $U(1)_R$ D-term. This is due to the
cancellation between the constant Fayet-Iliopoulos(FI) term
present in 4D gauged supergravity and the field-dependent FI term
coming from the internal gauge flux. On the other hand, the $S$
modulus and $M,\varphi$ are stabilized by the F-terms at the
perturbative regime. $Q'$ is also stabilized radiatively due to
the supersymmetric couplings to heavy fields without introducing
additional SUSY breaking sources. Then, the resulting F-terms for
$S$, $M,\varphi$ are negligible while the F-terms for $T,Q'$ are
$F^T\simeq 2m_{3/2}$, $F^{Q'}\simeq \sqrt{2}m_{3/2}$,
respectively, and the $U(1)_R$ D-term is $D_R\simeq
-\frac{m^2_{3/2}}{g_R}$. Consequently, because of the cancellation
between the moduli and hidden-brane F-terms\footnote{Due to the
absence of sequestering in 6D compactifications \cite{faleelud},
the contact term between hidden sector field $Q'$ and the visible
sector superfields makes the cancellation happen.}, scalar soft
masses are determined dominantly by the $U(1)_R$ D-term as
\begin{equation}
m^2_i\simeq r_ig_R D_R\simeq -r_i m^2_{3/2}\,.
\end{equation}
The nonzero soft masses for brane scalars proportional to the $R$-charges can
be also derived directly from the 6D action for the
non-supersymmetric flat brane solution with a small warping
\cite{warp6d}. The warping induced by unequal 3-brane tensions
makes the brane-localized and flux-induced masses uncancelled
\cite{U1R}. Therefore, as a small $U(1)_R$ D-term is generated below the compactification scale (or the 4D GUT scale),
it gives rise to a visible effect on the low-energy phenomenology by contributing 
to the initial SUSY spectrum at the GUT scale, unlike the conclusion of Castano et al in Ref.~\cite{Chamseddine95} where it was assumed that the $R$-symmetry is broken at the Planck scale and the $U(1)_R$ D-term vanishes.

\subsection{The $U(1)_R$ anomalies}

Assuming that the renormalizable Yukawa couplings are allowed for
quarks and leptons in the MSSM, the anomaly cancellation
conditions for the $U(1)_R$ determine the $R$-charges of
sfermions in terms of the squark doublet $R$-charge $\tilde q$
\cite{U1Rpheno} as follows, \bea
\tilde{l}&=&-3\tilde{q}-\frac{16}{3}, \quad \tilde{e}=-\frac{3}{7}\tilde{q} - \frac{26}{21}, \quad \tilde{u}=\frac{17}{7}\tilde{q}+\frac{18}{7},  \nonumber \\
\tilde{d}&=&-\frac{31}{7}\tilde{q}-\frac{46}{7}, \quad \tilde{h}_d= \frac{24}{7}\tilde{q}+\frac{60}{7}, \quad \tilde{h}_u=-\frac{24}{7}\tilde{q}-\frac{4}{7}. \label{scalarrcharge}
\eea
So, there is one parameter family of solutions to the consistent $R$-charges.
We note that the $R$-charge of a fermion differs from the one of scalar superpartner by one unit, as $l={\tilde l}-1$.
We assume that the pure $U(1)_R$ anomalies are cancelled by hidden fermions\footnote{See the $U(1)_R$ anomaly coefficients in the presence of hidden fermions with nonzero R-charges in Ref.~\cite{U1Rpheno}. There may be an additional R-symmetry breaking in the process of giving hidden fermions masses. If this occurs only in the hidden sector, there is no problem with interactions of the additional R-breaking fields and the MSSM fields. There might appear also light fermions and one of them could be the LSP. }.
On the other hand, it has been shown that nonzero $U(1)_R$-SM mixed anomalies, $C_a(a=1,2,3)$, can be cancelled by the variation of a Green-Schwarz term \cite{U1Rpheno}:
\be
{\cal L}_{GS}=({\rm Im}T)\sum_{a=1}^3 k_a\frac{1}{2}{\rm tr}(F^a {\tilde F}_a)
\ee
where the $U(1)_R$ gauge transform of ${\rm Im}T$ is $\delta_R ({\rm Im}T)=4g_R\Lambda_R$ and $k_a$ are related to the anomaly coefficients as $k_a=\frac{C_a}{16\pi^2 g_R}$ with $C_1=-15$ and $C_2=C_3=-9$.
Consequently, after a supersymmetric completion of the Green-Schwarz term, the gauge kinetic functions for the brane-localized SM gauge fields are modified to
\be
f_a=\frac{1}{g^2_{a,0}}+k_a T, \label{gaugekinterm}
\ee
where $g_{a,0}$ are the tree-level SM gauge couplings.
For unified tree-level gauge couplings with $g^2_{3,0}=g^2_{2,0}$ and $g^2_{1,0}=\frac{3}{5}g^2_{2,0}$ at the compactification scale,
$k_a=\frac{5}{3}k_2$ is consistent with the favorable choice of $\sin^2\theta_W=\frac{3}{8}$
at the compactification scale, as there is no exotics charged under the SM between the
unification scale and the electroweak scale.
However, given that the R-charges of the matter fields are non-universal in the same GUT multiplet, the matter multiplets should appear as split multiplets below the compactification scale as in orbifold GUT models.

\subsection{The $\mu$ term}

In the presence of the gauged $U(1)_R$,  the $\mu$ term is
forbidden at tree level by the consistent $R$-charge assignment in
(\ref{scalarrcharge}). Therefore, for the resolution of the $\mu$
problem, we introduce higher dimensional interactions with two
singlets, $Y,X$, in the superpotential\footnote{We note that other
dimension-5 singlet operators, $Y^2X^2$ and $Y^3X$, are
problematic because there is no minimum with nonzero singlet
VEV.}:
 \be W_\mu= \frac{h}{M_P}Y^2 H_u H_d +\frac{\kappa}{M_P} Y
X^3. \label{muterm}
 \ee
Here we note that there exists an
automatic Peccei-Quinn (PQ) symmetry\footnote{It is not possible
to write the self-interaction for the singlet in the
superpotential to break PQ symmetry explicitly.} which provides
the axion solution of the strong CP problem \cite{Kim87}. 
As shown in Appendix A, in the presence of weak-scale soft mass
terms, the dimension-5 interaction \footnote{If the $\mu$ term comes from a renormalizable singlet
interaction, a necessary small singlet VEV would lead to a
dangerous axion due to low PQ symmetry breaking scale.} between singlets gives rise to
intermediate-scale singlet VEVs so that we can get a weak-scale
$\mu$ term.  Note that
this realizes the idea of Ref.~\cite{Kim84} by imposing a
fundamental $U(1)_R$ symmetry. In the same appendix, we have shown the mass spectrum of the singlet sector after minimizing the singlet potential. It turns out that the masses of axino and saxion partners in $X,Y$ singlets are heavier than the gravitino mass. 
The details for the axion property
are explained in Appendix B. 
Even with an explicit $PQ$-breaking term and after the $R$-symmetry breakdown
for nonzero singlet VEVs, the PQ symmetry breaking is small enough
for maintaining the axion solution to the strong CP problem. 
The details on this aspect are shown in Appendix C.

\subsection{Neutrino masses}

In this section, we introduce  right-handed neutrinos which can
have Majorana or Dirac masses. It further constrains  the allowed
$R$-charges of the MSSM sector, consequently determining the
scalar soft masses via the $U(1)_R$ mediation. We first consider
the Majorana neutrino case with intermediate or TeV-scale Majorana
mass for the RH neutrino. Then, we go on to discuss the Dirac
neutrino case with vanishing Majorana mass for the RH neutrino.

\subsubsection{Majorana neutrino case}

We consider the neutrino mass term in the superpotential  with
right-handed neutrino $N$ as follows, \be W_\nu= \lambda_\nu L H_u
N+\frac{\lambda_N}{2M_P^{n-1}} X^n NN. \label{Wmajorana} \ee The
standard high-scale see-saw mechanism is applied for $n=1$ while
the TeV-scale see-saw mechanism must be used for $n=2$. Then, for
the $R$-charge of the $Y$ singlet, $r_Y=-3$, we obtain the
$R$-charges of the other singlets as $r_X=\frac{5}{3}$ and
$r_N=1-\frac{5n}{6}$. Then, the $R$-charge of the squark doublet,
that determines all the other $R$-charges through
Eq.~(\ref{scalarrcharge}), is determined to be \be {\tilde
q}=-\frac{29}{27}-\frac{7n}{54}.\label{squarkrcharge} \ee Thus, we
obtain the doublet squark $R$-charge to be ${\tilde
q}=-\frac{65}{54}$ for $n=1$ and ${\tilde q}=-\frac{4}{3}$ for
$n=2$. Note that the $PQ$-charges can be assigned in an
appropriate way and thus the axion solution to the strong CP
problem persists even after introducing right-handed neutrinos.
For $n=2$, the $R$-charges and $PQ$-charges are shown in Table 1.

\begin{table}[!h]
\begin{center}
\begin{tabular}{|c|c|c|c|c|c|c|c|c|c|c|}
\hline
   & Q & U  & D  & L & E & $H_u$ & $H_d$  & Y  & X  & N   \\ \hline
$U(1)_R$ & $-\frac{4}{3}$ & $-\frac{2}{3}$ & $-\frac{2}{3}$  & $-\frac{4}{3}$  & $-\frac{2}{3}$ & 4 & 4 & $-3$ & $\frac{5}{3}$ & $-\frac{2}{3}$  \\ \hline
$U(1)_{PQ}$ & $-3$ & 0 & 0 & $-2$ & $-1$ & 3 & 3 & $-3$ & $1$ & $-1$ \\
\hline
\end{tabular}
\end{center}
\caption{ R-charges and PQ-charges for the Majorana neutrino case with $n=2$}\label{charges1}
\end{table}
The PQ symmetry is nothing but a global $R$-symmetry with new $R$-charges ${\tilde r}_i$ given by the shifted ones from the local $R$-charges, ${\tilde r}_i=r_i+q_i$ with $q_i$ being the PQ charges.
As shown in the Appendix C, the PQ symmetry is broken explicitly by higher order $U(1)_R$-invariant terms in the superpotential. The same is true of the Dirac neutrino case.

\subsubsection{Dirac neutrino case}

We note that in the absence of the Majorana neutrino mass term, it
is possible in our framework to realize the tiny Dirac neutrino
Yukawa coupling for neutrino masses in the superpotential.

One possibility (Type I) is to take the following Dirac neutrino
Yukawa coupling, \be W_\nu= \frac{\lambda_\nu}{M^2_P} XY L H_u
N.\label{WdiracI} \ee Since the $R$-charge of the right-handed
(RH) sneutrino becomes \be r_N=\frac{45}{7}{\tilde
q}+\frac{194}{21}, \label{rnrcharge} \ee the $R$-charges of all
fields except $X,Y$ are determined in terms of the doublet squark
$R$-charge. We note that the doublet squark $R$-charge is not
determined unlike the Majorana neutrino case. For this type of
Dirac neutrino mass, taking into account the condition on $\tilde
q$ for getting positive soft squared masses of squarks, sleptons
and RH sneutrino, we give an example with rational $R$-charges in
Table 2.
\begin{table}[!h]
\begin{center}
\begin{tabular}{|c|c|c|c|c|c|c|c|c|c|c|}
\hline
   & Q & U  & D  & L & E & $H_u$ & $H_d$  & Y  & X  & N   \\ \hline
$U(1)_R$ & $-\frac{13}{9}$ & $-\frac{59}{63}$ & $-\frac{11}{63}$  & $-1$  & $-\frac{13}{21}$ & $\frac{92}{21}$ &
$\frac{76}{21}$ & $-3$ & $\frac{5}{3}$ & $-\frac{1}{21}$  \\ \hline
$U(1)_{PQ}$ & $-3$ & 0 & 0 & $-2$ & $-1$ & 3 & 3 & $-3$ & $1$ & $1$ \\
\hline
\end{tabular}
\end{center}
\caption{ R-charges and PQ-charges for the Dirac neutrino case}\label{charges2}
\end{table}

The other possibility (Type II) is to take the following
superpotential; \be W_\nu=\frac{\lambda_\nu}{M^2_P} X^2 L H_u
N.\label{WdiracII} \ee Then, the $R$-charge of the RH sneutrino is
\be r_N=\frac{45}{7}{\tilde q}+\frac{32}{7}. \ee We note that the
$Y^2LH_u N$ coupling would lead to the $R$-charge of the RH
sneutrino as $r_N=\frac{45}{7}{\tilde q}+\frac{292}{21}$, giving
rise to a tachyonic RH sneutrino for the allowed range of $\tilde
q$.

In either case without tachyonic RH sneutrino, a necessary tiny
Dirac Yukawa coupling can be generated when the singlets get
intermediate-scale VEVs as shown in Appendix A. For instance, in
the former case, we obtain the neutrino mass as
\be
m_\nu= y_\nu v \sin\beta \simeq
0.01 \ {\rm eV}\,, \label{mnu}
\ee
where $y_\nu \equiv \frac{\lambda_\nu}{M^2_P}\langle XY\rangle$.
 Thus, plugging the singlet VEVs
~(\ref{singletvevs}) in the above, we require
$\frac{\lambda_\nu}{\kappa}\sim 10^{4}$ for
$\frac{m_{3/2}}{M_P}\sim 10^{-16}$. In Type II case, we would
need $\frac{\lambda_\nu}{\kappa}\sim 10^{3}$ for the same
gravitino mass.

\section{Initial soft mass parameters}

Let us first summarize  the $U(1)_R$-mediated SUSY breaking in the
MSSM sector. Then, we also determine the soft mass parameters for
the singlet sector, that is responsible for generating the $\mu$
term as well as the neutrino masses.

For a scalar field with $R$-charge $r_i$, the $U(1)_R$ D-term
determines the scalar soft mass \cite{U1Rpheno} as $m^2_i=-r_i
m^2_{3/2}$. Thus, from the $R$-charges of the MSSM fields in
Eq.~(\ref{scalarrcharge}), the scalar soft masses are given by
\bea m^2_{\tilde q}&=&-{\tilde q}m^2_{3/2}, \quad
m^2_{\tilde{l}}=\Big(3\tilde{q}+\frac{16}{3}\Big)m^2_{3/2},
\nonumber \\   m^2_{\tilde{e}}&=&\Big(\frac{3}{7}\tilde{q} +
\frac{26}{21}\Big)m^2_{3/2}, \,
m^2_{\tilde{u}}=-\Big(\frac{17}{7}\tilde{q}+\frac{18}{7}\Big)m^2_{3/2},
\,
m^2_{\tilde{d}}=\Big(\frac{31}{7}\tilde{q}+\frac{46}{7}\Big)m^2_{3/2}, \nonumber \\
 m^2_{\tilde{h}_d}&=& -\Big(\frac{24}{7}\tilde{q}+\frac{60}{7}\Big)m^2_{3/2},
  \quad m^2_{\tilde{h}_u}=\Big(\frac{24}{7}\tilde{q}+\frac{4}{7}\Big)m^2_{3/2}. \label{scalarmass}
\eea
 Note that   all squarks and
leptons squared masses are positive when the doublet squark
$R$-charge lies in the range $-\frac{46}{31}<{\tilde
q}<-\frac{18}{17}$. In this $R$-charge range, the soft mass
squareds of the scalar Higgs doublets are negative. The
corresponding trilinear soft terms for Yukawa couplings are
universal as \be A_{ijk}=-2m_{3/2} \ee for all $i,j,k$.

From the modified gauge kinetic term, Eq.~(\ref{gaugekinterm}),
in the presence of the nonzero F-term of the $T$-modulus, the
gaugino masses for the SM gauge group are given by the $U(1)_R$-SM
mixed anomalies and are universal at the GUT scale:
 \be M_a=k_a
g^2_a F^T\simeq -\frac{9}{16\pi^2 g_R} m_{3/2}, \quad a=1,2,3,
 \ee
where the used relations are $16\pi^2 g_R k_a g^2_a=-9 g^2_{\rm
GUT}\simeq -\frac{9}{2}$ and $F^T\simeq 2m_{3/2}$. The gaugino
masses can be larger or smaller compared to the gravitino mass
depending on the $U(1)_R$ gauge coupling (e.g. $|M_a|\gtrsim
m_{3/2}$ for $g_R\lesssim \frac{9}{16\pi^2}$). In particular, for
a small $U(1)_R$ gauge coupling as required for a large gaugino
mass, the Green-Schwarz terms with large $k_a$ give a large
negative contribution to the SM gauge kinetic terms. In this case,
to get the unified value of the gauge couplings, $g^2_{\rm
GUT}\simeq \frac{1}{2}$, we need to cancel the large contribution
of the Green-Schwarz term by considering small tree-level SM gauge
couplings. Henceforth we treat the universal gaugino mass
$M_{1/2}$ to be a free parameter. We will see that $M_{1/2} \gg
m_{3/2}$ is required for a proper electroweak symmetry breaking.

\subsection{Majorana neutrino case}

For the Majorana neutrino case, the soft mass terms for singlets are
\bea
{\cal L}_{\rm soft} &\supset& -m^2_X |{ X}|^2 -m^2_Y|{ Y}|^2 -m^2_N |{ N}|^2  \\
&& -\frac{{ h}}{M_P}A_h { Y}^2 H_u { H}_d -\frac{
\kappa}{M_P}A_\kappa { Y}{ X}^3 -{\lambda}_\nu A_\nu{ L}{ H}_u{
N}-\frac{{\lambda}_N}{2M^{n-1}_P}A_N{ X}^n{ N}{ N}+{\rm c.c.} \,.
\nonumber  \label{singletsoftmass}
 \eea
 The soft mass parameters for the singlet sector are
determined in the $U(1)_R$ mediation as follows: \be
m^2_X=-\frac{5}{3}m^2_{3/2}, \ \ m^2_Y=3 m^2_{3/2}, \ \
m^2_N=\Big(-1+\frac{5n}{6}\Big)m^2_{3/2}\,.\label{singletmass} \ee
The $A$ terms in the neutrino sector also follow the relation $A
\simeq -2 m_{3/2}$ as shown below;
 \bea
 A_h&=&-F^I\partial_I
\ln\Big(\frac{h}{CY^2_Y Y_{H_u}
Y_{H_d}}\Big)=\frac{F^C}{C_0}+4\Big(\frac{F^S}{6s}-\frac{F^T}{3t}\Big)
\simeq -2m_{3/2},\nonumber \label{mutril} \\
A_\kappa&=&-F^I\partial_I \ln\Big(\frac{\kappa}{C Y_Y Y^3_X}\Big)\simeq -2m_{3/2},
 \nonumber\label{selftril}\\
A_\nu&=&-F^I\partial_I \ln\Big(\frac{\lambda_\nu}{ Y_L Y_{H_u}Y_N}\Big)
=3\Big(\frac{F^S}{6s}-\frac{F^T}{3t}\Big)\simeq -2m_{3/2},
\nonumber\label{diracnutril} \\
A_N&=&-F^I\partial_I \ln\Big(\frac{\lambda_N}{C^{n-1} Y^n_X
Y^2_N}\Big)
=(n-1)\frac{F^C}{C_0}+(n+2)\Big(\frac{F^S}{6s}-\frac{F^T}{3t}\Big)
\simeq -2 m_{3/2} \nonumber
 \eea
where $\frac{F^C}{C_0}\simeq \frac{2}{3}m_{3/2}$ and $F^S\ll
F^T\simeq 2m_{3/2}$. Here, $Y_i$'s are defined from the expansion
of the superconformal factor, $\Omega=-3 e^{-K/3}$: $\Omega\simeq -3
e^{-K_0/3}+Y_i Q_i^\dagger Q_i$ where $Q_i$ are all the
brane-localized chiral superfields, $K_0$ is independent of the
brane fields, and
$Y_i=\Big(\frac{1}{2}(S+S^\dagger)\Big)^{1/3}\Big(\frac{1}{2}(T+T^\dagger)-Q^{\prime
\dagger}Q'-\varphi^\dagger\varphi\Big)^{-2/3}.$
 In the presence of
nonzero singlet VEVs, we obtain $\mu,B\mu$ terms as follows, \bea
\mu&=&\frac{h}{M_P}\langle Y^2\rangle, \label{mu1}\\
\quad B \mu &=& A_h \mu +\frac{2h\kappa^*}{M^2_P}\langle Y X^{*3}\rangle=\mu\Big(A_h
+\frac{2\kappa^*}{M_P}\langle Y^{-1}X^{*3}\rangle\Big). \label{B1}
 \eea
After the singlet VEVs (\ref{singletvevs}) are inserted in the
above, we find that \bea
\mu&\simeq & 0.0272 \,\frac{h}{\kappa} \,m_{3/2},  \label{muterm1}\\
B&\simeq & 7.49 \,m_{3/2}. \label{Bterm}
 \eea
Moreover, the RH neutrino masses are also determined as follows,
\bea
M_N&=&\frac{\lambda_N}{M^{n-1}_P}\langle X^n\rangle, \label{MN0} \\
 B_N M_N &=& A_N M_N + \frac{3n\lambda_N \kappa^*}{M^n_P}
  \langle Y^{*} X^{*2} X^{n-1}\rangle \nonumber \\
  &=&M_N\Big(A_N+\frac{3n\kappa^*}{M_P}\langle Y^{*}X^{*2}X^{-1}\rangle\Big). \label{BN0}
\eea
 In the $n=2$ case, from Eq.~(\ref{scalarmass}) with
$R$-charges given in Table 1, the MSSM scalar soft masses at the
GUT scale are determined as follows, \bea
m^2_{\tilde q}&=&m^2_{\tilde l}=\frac{4}{3}m^2_{3/2}, \quad m^2_{\tilde u}=m^2_{\tilde d}=m^2_{\tilde e}=\frac{2}{3}m^2_{3/2},  \nonumber \\
m^2_{H_u}&=&m^2_{H_d}=-4m^2_{3/2}. \label{mssmscalarmass}
\eea
In this case, from Eqs.~(\ref{MN0}) and (\ref{BN0}), the RH neutrino masses are
\bea
M_N&\simeq& 0.849\,\frac{\lambda_N}{\kappa} \,m_{3/2}, \label{MN}\\
B_N &\simeq& -1.09 m_{3/2} \label{BN}.
 \eea
Then, the mass eigenvalues of the RH sneutrino are \dis{
m^2_{\tilde{N}_\pm}&=M_N^2+m_N^2\pm|B_N|M_N\\
&\simeq
m^2_{3/2}\left[\left(0.849\frac{\lambda_N}{\kappa}\right)^2+\frac23\pm
  \left(0.849\frac{\lambda_N}{\kappa}\right)\right]\ge \frac{5}{12}m^2_{3/2} .\label{majoranasneutrino}
}

\subsection{Dirac neutrino case}

In the Dirac neutrino case with $XYLH_u N$, the soft mass terms for singlets are
\bea
{\cal L}_{\rm soft} &\supset& -m^2_X |{ X}|^2 -m^2_Y|{ Y}|^2 -m^2_N |{ N}|^2 \nonumber \\
&& -\frac{{ h}}{M_P}A_h { Y}^2 { H}_u { H}_d -\frac{
\kappa}{M_P}A_\kappa { Y}{ X}^{3}
-\frac{{\lambda}_\nu}{M^2_P}A_\nu { X}{ Y} { L}{ H}_u{ N}+{\rm
c.c.}
\eea
 The scalar soft masses for the $X,Y$ singlets and the
trilinear couplings corresponding to the $\mu$ term and the
singlet interaction are the same as in the Majorana neutrino case,
so, after $X,Y$ singlets get VEVs, the $\mu$ and $B\mu$ terms are
given by Eqs.~(\ref{muterm}) and (\ref{Bterm}), respectively.

As the $R$-charges of all fields are determined in terms of the
squark $R$-charge, so are the scalar soft masses in the $U(1)_R$
mediation. Then, the RH sneutrino scalar soft mass is given by
$m^2_N=-\Big(\frac{45}{7}{\tilde q}+\frac{194}{21}\Big)
m^2_{3/2}$. For the doublet squark $R$-charge,
$-\frac{46}{31}<{\tilde q}<-\frac{194}{135}$, not only all the
squarks and sleptons but also the RH sneutrino have positive
scalar squared soft masses\footnote{For the Dirac Yukawa coupling
$X^2LH_uN$, the RH sneutrino mass is given by
$2.24m^2_{3/2}<m^2_N<4.97m^2_{3/2}$.} as
\bea 0<m^2_N<0.301\,
m^2_{3/2}\label{mNDiracI}.
\eea
 We note that, in this region of
$|{\tilde q}|$, the relic density coming from neutralino as the
LSP tends to be too large \cite{U1Rpheno}. So,  it is natural to
take the RH sneutrino or gravitino as a dark matter candidate. The
trilinear coupling for the Dirac neutrino Yukawa coupling is given
by $$ A_\nu
=-F^I\partial_I\ln\bigg(\frac{\lambda_\nu}{C^2Y_XY_YY_LY_{H_u}Y_N}\bigg)=2\frac{F^C}{C_0}
+5\Big(\frac{F^S}{6s}-\frac{F^T}{3t}\Big)\simeq -2m_{3/2} $$
 following again the relation $A_{ijk} \simeq -2m_{3/2}$.

\section{Low energy spectrum and dark matter}

In this section, we consider the constraints on the SUSY spectrum
coming from the EWSB conditions. Even after the $R$-symmetry
breakdown, the $R$-parity is a good symmetry at the perturbative
level as the $R$-parity violating terms appear at sufficiently
higher orders as shown in Appendix D. Depending on the nature of
neutrino masses, we take either gravitino or RH sneutrino to be a
dark matter candidate. We discuss  the dark matter relic density
and the BBN constraints on a late decaying NLSP in either dark
matter scenario.

\subsection{The EWSB condition and the SUSY spectrum}

The Higgs mass terms contributing to the Higgs potential are given
by \be
V_{h,mass}=(|\mu|^2+m^2_{H_u})|H_u|^2+(|\mu|^2+m^2_{H_d})|H_d|^2+(B\mu
H_u H_d+{\rm c.c.}) \ee In order to achieve electroweak symmetry
breaking,  the following conditions at the weak scale must
be fulfilled:
\bea |B\mu|^2>
(|\mu|^2+m^2_{H_u})(|\mu|^2+m^2_{H_d}), \eea \bea
2|\mu|^2+m^2_{H_u}+m^2_{H_d} -2|B\mu|>0.
\eea
 Then, the
minimization conditions for the Higgs potential impose the  following
conditions;
\bea
\sin(2\beta)&=&\frac{2|B\mu|}{m^2_{H_u}+m^2_{H_d}+2|\mu|^2}, \label{ewsb1}\\
|\mu|^2&=&\frac{m^2_{H_d}-m^2_{H_u}\tan^2\beta}{\tan^2\beta
-1}-\frac{M^2_Z}{2}. \label{ewsb2}
\eea
 The above EWSB conditions require that the $\mu$-term
 and Higgs scalar soft masses must
be large at the EWSB scale to be compatible with the large
$B$-term  as compared to gravitino or scalar soft masses at the
GUT scale in \eq{Bterm}. Therefore, the necessary large loop
corrections to the Higgs scalar soft masses can be obtained for
the gaugino mass which is much larger than the gravitino mass.

When $M_{1/2}\gg \mgravitino$,
after RGE running from GUT scale to EWSB scale, the soft terms at the EWSB scale
becomes of the order the gaugino mass at GUT scale while $B$ and $\mu$-terms
do not change much.
For $\tgbeta \gtrsim 1$, considering $|\mu|^2\simeq-m^2_{H_u}\sim M_{1/2}^2$ from
\eq{ewsb2}, we find roughly
\dis{
B\simeq 7.5\mgravitino\sim \frac{|\mu|}{\tgbeta}\sim \frac{M_{1/2}}{\tgbeta}.
}
at the EWSB scale.
This is the common feature of this $U(1)_R$ gauged model.
For the correct magnitude for $\mu$-term we need a large ratio between $h$
and $\kappa$ as $h/\kappa\sim 10^3$ from \eq{muterm}.

Since gaugino mass is much larger than scalar soft masses at GUT scale,
at low energy, all the masses of the SUSY particles are of the order of the
gaugino
mass and the lightest ordinary supersymmetric particle becomes
lighter stau $\stau$.
Therefore the possible Dark Matter(DM) candidate must be
gravitino or sneutrino which is outside of the MSSM sector.
We note that the axino and saxion partners of the $X,Y$ singlets are heavier than gravitino
or sneutrino so they cannot be LSP.
In the following sections we consider the corresponding DM candidate
for each model introduced in the previous section.

\subsection{Majorana neutrino case}

First, we consider the Majorana neutrino with $n=1$ in
\eq{Wmajorana}, the mass of the scalar RH neutrino mass is
$M_{N}\simeq\lambda_N\VEV{X}\sim 10^{10-12} {\rm GeV}$. Thus, gravitino, as
the LSP, is the only dark matter candidate while stau is the NLSP.
In this case, there are two sources for the relic
density of gravitino DM: non-thermal production from the decay of
stau NLSP and thermal production from the thermal scattering after
reheating. We will not consider the thermal production which
depends on the reheating temperature.

When NLSP decays after freezeout, the non-thermal production of LSP dark matter is determined by
 \begin{equation}
  \Omega_{DM}h^2 =  { m_{\rm DM}
  \over m_{\rm NLSP} } \Omega_{\rm NLSP}h^2. \label{relicdensity}
 \end{equation}
 The abundance of stau from the thermal
freeze-out is \cite{aim}
 \dis{ \Omega_\stau h^2\simeq
 0.2\bfrac{\mstau}{1 \tev}^2.\label{oh2stau} }
For gravitino LSP and  stau NLSP which decay via
${\tilde\tau}_1\to \tau + {\tilde G}$, from
Eqs.~(\ref{relicdensity}) and~(\ref{oh2stau}), the non-thermal
production of gravitino is
 \dis{ \Omega_\gravitino h^2\simeq 0.02
 \bfrac{\mgravitino}{100\gev}\bfrac{\mstau}{1\tev}.\label{oh2Gn1} }
For a correct relic density, we need a stau mass of TeV scale which is consistent with the EWSB conditions
in our scenario.

However the decay products of a long-lived decaying particle can
be problematic with respect to the standard BBN. When a long-lived particle is negatively charged, it can make a bound
state with nuclei and even worse the situation (CBBN)~\cite{CBBN}.
To avoid the CBBN constraint, the lifetime of stau must be less than
$5\times10^3\sec$ or $Y_{\stau}\equiv n_{\stau}/s\lesssim
10^{-15}$ for longer
lifetime~\cite{Hamaguchi:2007mp,Kawasaki:2007xb,Pradler:2007is,Jedamzik:2007qk,Jedamzik:2009uy}.
Such a small abundance of stau requires unusual
situations~\cite{Ratz:2008qh,Pradler:2008qc,Bailly:2009pe,Boubekeur:2010nt}.
As stau decays dominantly to gravitino and tau
lepton, the lifetime of stau is given by\dis{ \tau\left(\stau\rightarrow \gravitino + \tau
\right)
&\simeq 48\pi \Mp^2\frac{m^2_{3/2}}{m^5_{{\tilde\tau}_1}}\\
&\simeq 1.8\times10^3\sec \bfrac{\mgravitino}{100\gev}^2\bfrac{1\tev}{\mstau}^5. \label{staulife}
}
The CBBN constraint is automatically satisfied for $m_\stau \gtrsim 2$ TeV
in the region of producing the right relic
density of gravitino from stau decay according to \eq{oh2Gn1}.
For example, for $\tgbeta=10$, our model with correct EWSB predicts
$\mgravitino\simeq 200 \gev$ and $\mstau=2.6 \tev$,
which leads to the lifetime of stau around 100 sec.

In the case of Majorana neutrino with $n=2$ in \eq{Wmajorana}, the
mass of RH sneutrino is determined by
Eq.~(\ref{majoranasneutrino}). If we consider the small magnitude
of $\kappa\sim {\mathcal O}(10^{-3})$ with $\lambda_N\sim
{\mathcal O}(1)$, then $m_{\tilde{N}_\pm}\simeq
(0.85\lambda_N/\kappa)\mgravitino \sim 10^3\mgravitino$ and
becomes much heavier than the other SUSY particles as well as
gravitino. This gives gravitino DM as in the $n=1$ case.
For $\lambda_N\sim\kappa$, we may obtain $m_{\tilde{N}_\pm}\sim
\mgravitino$ so two scenarios are possible: RH
sneutrino is LSP and gravitino is NLSP and vice versa. In both cases,
however, the thermal production of the RH sneutrino LSP
would highly overclose the Universe \cite{aim}, and thus it is excluded.
\begin{figure}[!t]
  \begin{center}
  \begin{tabular}{c c}
   \includegraphics[width=0.5\textwidth]{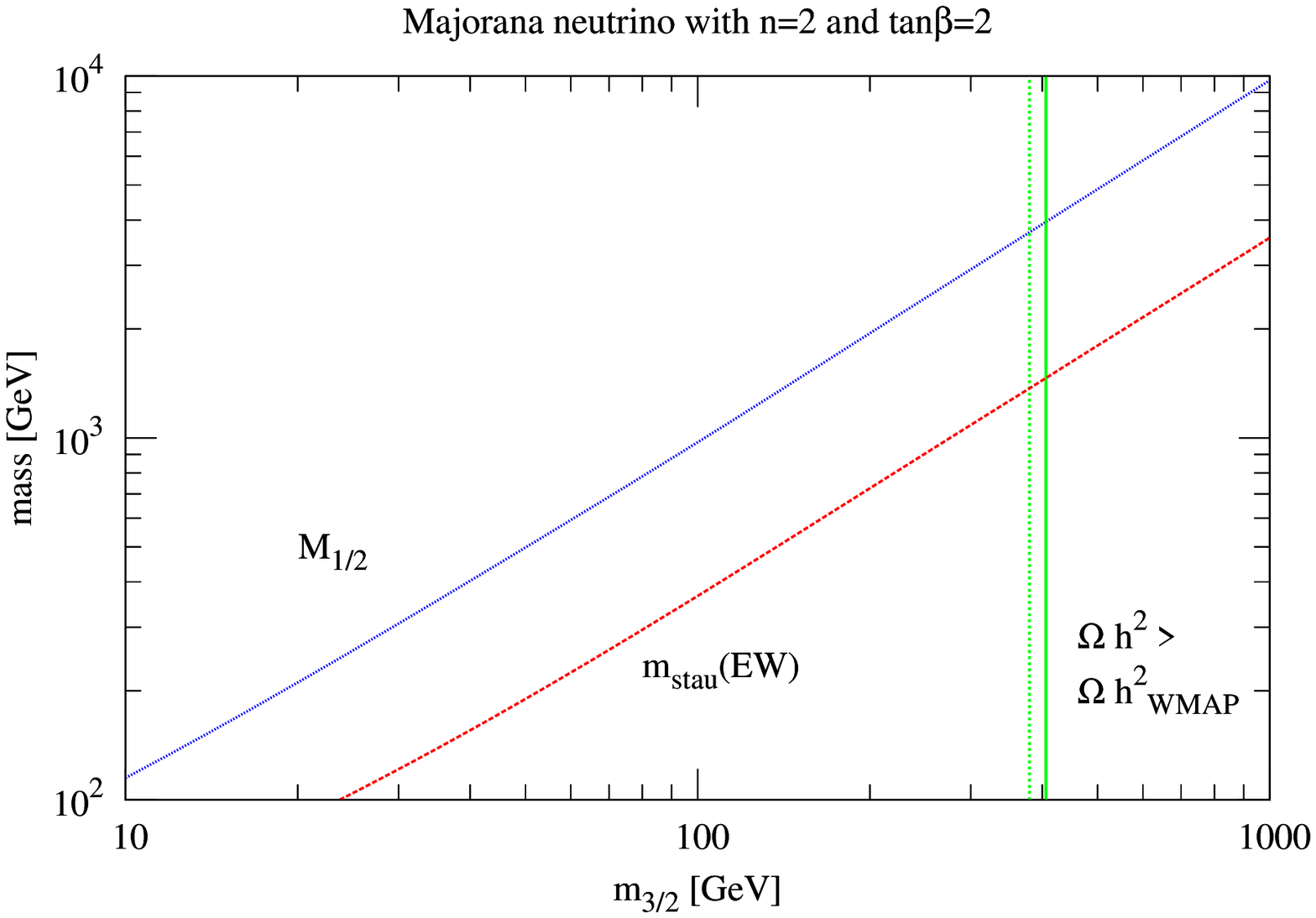}
&
   \includegraphics[width=0.5\textwidth]{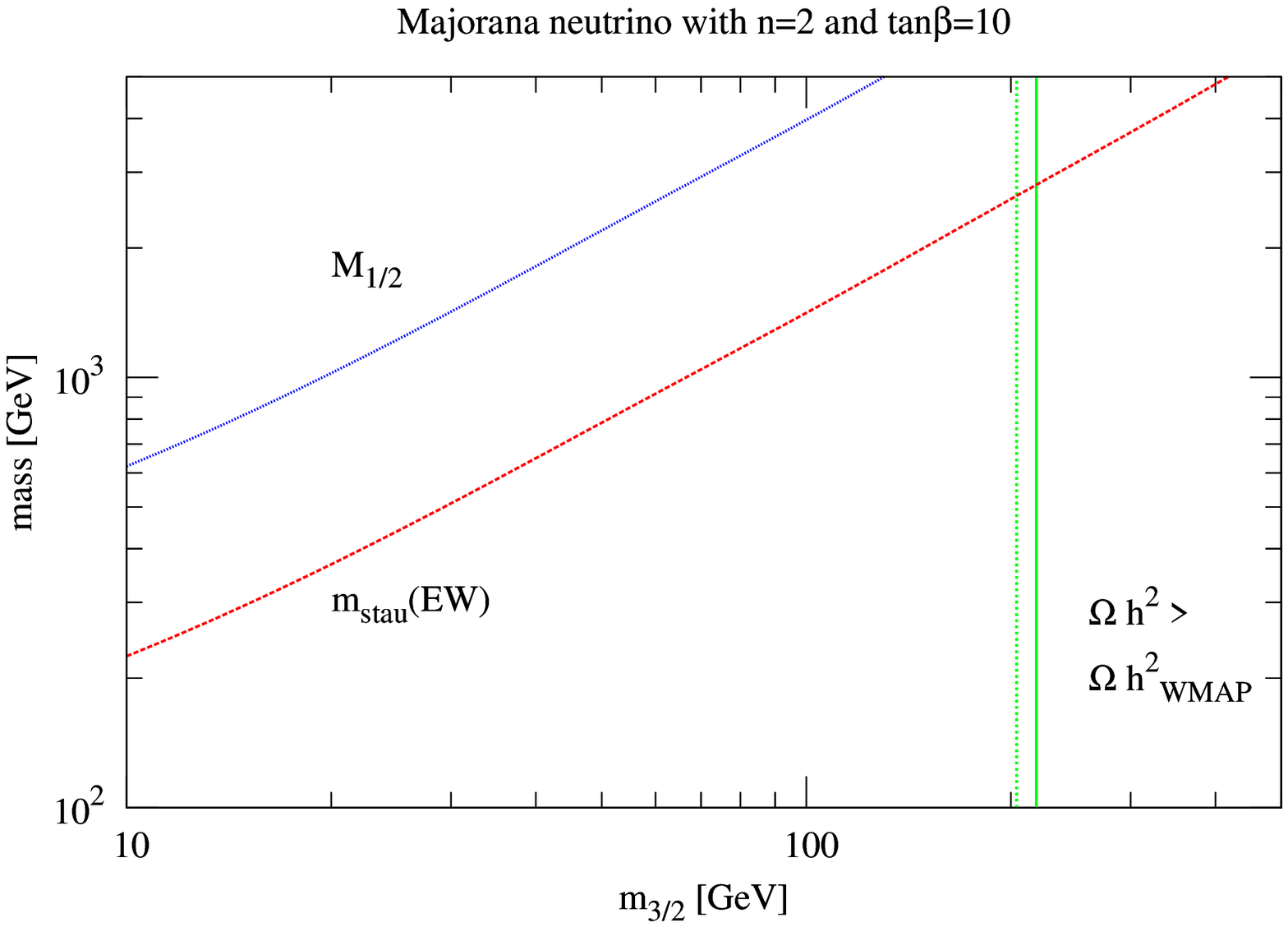}
   \end{tabular}
  \end{center}
 \caption{Plot of gaugino mass at GUT scale (upper blue line) and stau NLSP at
   EWSB (lower red line)
for a given gravitino mass with correct EWSB condition in the case
of Majorana neutrino with $n=2$. Between the vertical green lines, the
non-thermally produced gravitino from stau decay satisfies the
correct dark matter abundance by WMAP. }
\label{plot_ewsb_op2}
\end{figure}

In Figure~\ref{plot_ewsb_op2}, we show the gaugino mass $M_{1/2}$
at the GUT scale  and the stau mass at the EWSB scale vs the gravitino mass
after taking into account the correct EWSB. The green lines show the
region where the gravitino non-thermal production from stau decay
is within the range of cold dark matter from WMAP 7-year
data~\cite{Komatsu:2010fb}, $0.105 < \Omega_{DM} h^2 < 0.119$. The
region right to the solid green line, where  $ \Omega_{DM} h^2\
> 0.119$, is excluded. In Figure 2, we show the region of correct relic density in the plane of $\tan\beta$
and $m_{3/2}$.
\begin{figure}[!t]
  \begin{center}
  \begin{tabular}{c c}
   \includegraphics[width=0.5\textwidth]{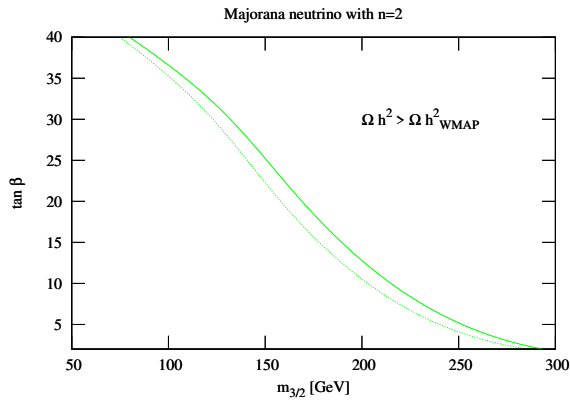}
   \end{tabular}
  \end{center}
 \caption{The contour plot correct relic density of gravitino from non-thermal
   production in the plane of $\tgbeta$ and gravitino mass in the case of Majorana
neutrino with $n=2$.
}
\label{plot_tanb_mG_op2}
\end{figure}

Gravitino can be produced thermally, depending on the reheating
temperature after inflation. In the region left to the green
lines, the thermal production  of gravitino is required for the
correct relic density of gravitino.

\subsection{Dirac neutrino case}

First, we consider Dirac neutrino Type I with the
superpotential~\eq{WdiracI}. Since the renormalization group
evolution of the (Dirac) RH sneutrino mass $m_{\tilde{N}}$
is negligible due to the smallness of
Yukawa coupling, the RH sneutrino mass at EWSB is smaller than the
gravitino mass from \eq{mNDiracI}. Therefore the RH sneutrino is
the LSP Dark Matter, and  gravitino is the NLSP.

\begin{figure}[!t]
  \begin{center}
  \begin{tabular}{c c}
   \includegraphics[width=0.5\textwidth]{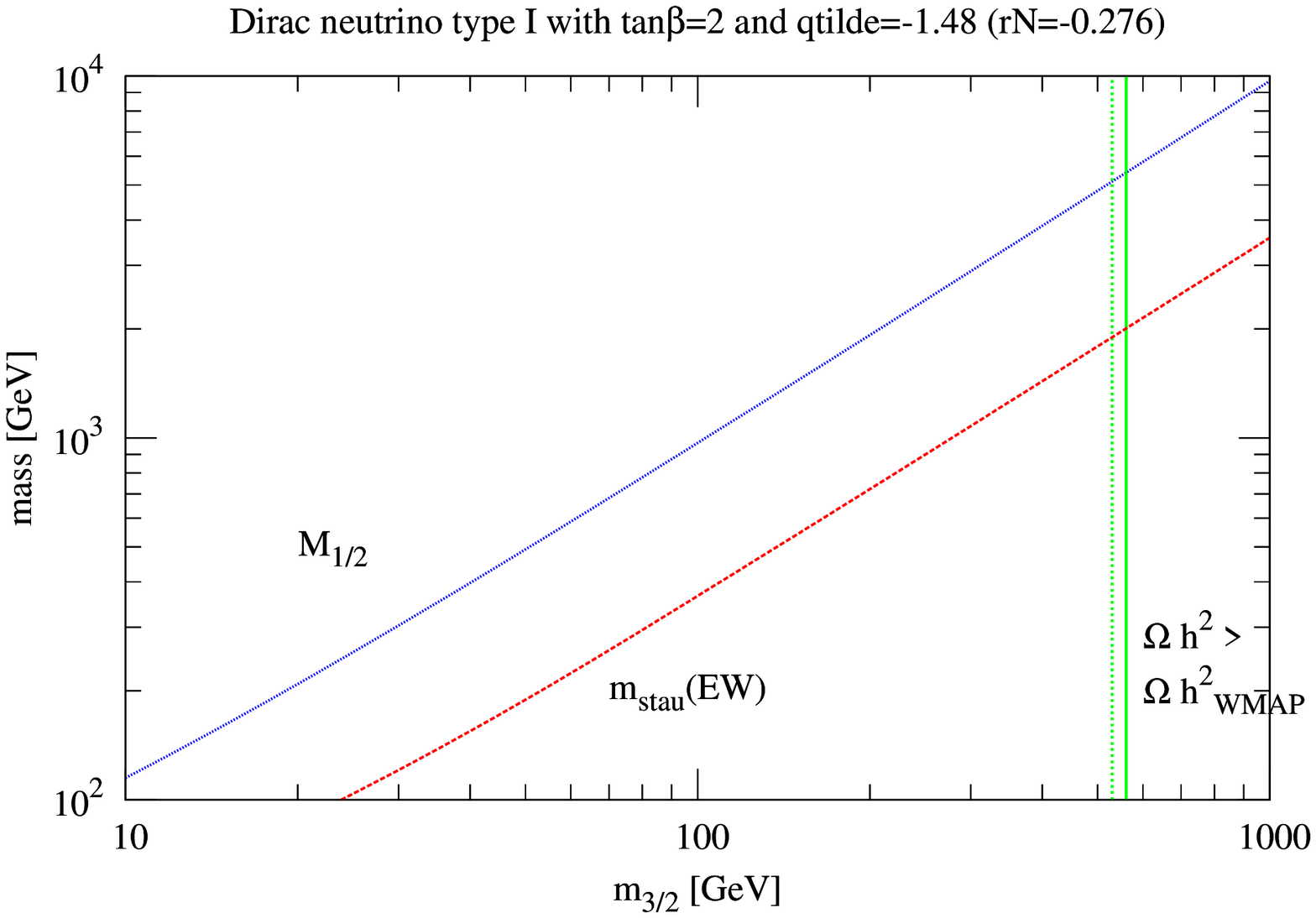}
&
   \includegraphics[width=0.5\textwidth]{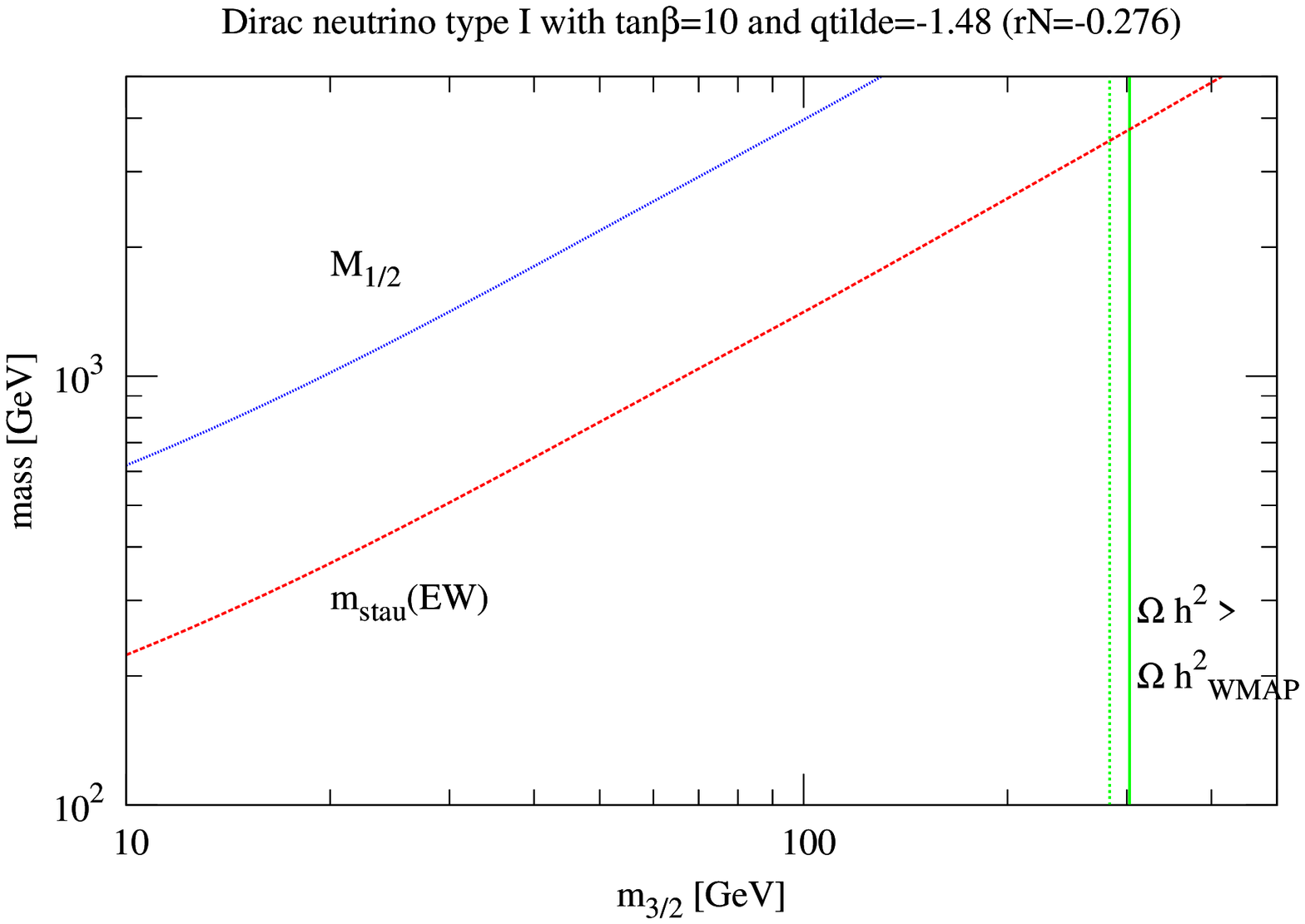}
   \end{tabular}
  \end{center}
 \caption{The same as Figure~\ref{plot_ewsb_op2} but in the case of Dirac
neutrino type I.}
\label{plot_ewsb_op3}
\end{figure}

\begin{figure}[!t]
  \begin{center}
  \begin{tabular}{c c}
   \includegraphics[width=0.5\textwidth]{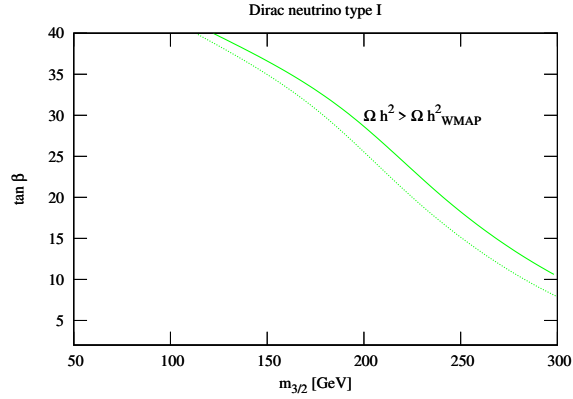}
   \end{tabular}
  \end{center}
 \caption{The same as Figure~\ref{plot_tanb_mG_op2} but in the case of Dirac
neutrino type I.
}
\label{plot_tanb_mG_op3}
\end{figure}

Let us note that the thermal production of the RH sneutrino can be
obtained from the decay of supersymmetric particles but it is
suppressed by the small Dirac neutrino Yukawa coupling
$y_\nu \sim 10^{-13}$.
If there is no enhancement factor due
to a small mass difference between left-handed and RH sneutrinos
or degenerate neutrino masses (requiring large $y_\nu$),
one typically gets the relic density
of the RH sneutrino from the thermal production as
$\Omega_{\tilde{N}_\pm} h^2 < {\cal O}(10^{-3})$ \cite{aim}. Thus,
the main contribution to the RH sneutrino DM can come from the
non-thermal production due to the decay of the NLSP. Using
Eqs.~(\ref{relicdensity}) and~(\ref{oh2stau}) for RH sneutrino
LSP, the relic density of the RH sneutrino produced non-thermally
from stau decay is
\dis{ \Omega_{{\tilde N}_\pm}h^2 =
\frac{m_{{\tilde N}_\pm}}{\mstau}\Omega_\stau h^2 \simeq
0.02\bfrac{m_{{\tilde N}_\pm}}{100\gev}\bfrac{\mstau}{\tev}, }
 which includes the RH sneutrino produced from stau decay to
gravitino and gravitino decay to RH sneutrino. The region which
gives the correct relic density for DM is shown in the
Figure~\ref{plot_ewsb_op3} for fixed $\tgbeta=2,10$ and in the
Figure~\ref{plot_tanb_mG_op3} on the $\tgbeta$ and $m_{3/2}$
plane. Here we used $r_N= -0.276$ so that the mass of RH sneutrino
is $m_{\tilde{N}_\pm}= 0.52\mgravitino$. With smaller $r_N$ the
green lines in the Figures move to the right direction
correspondingly.

\begin{figure}[!t]
  \begin{center}
  \begin{tabular}{c c}
   \includegraphics[width=0.5\textwidth]{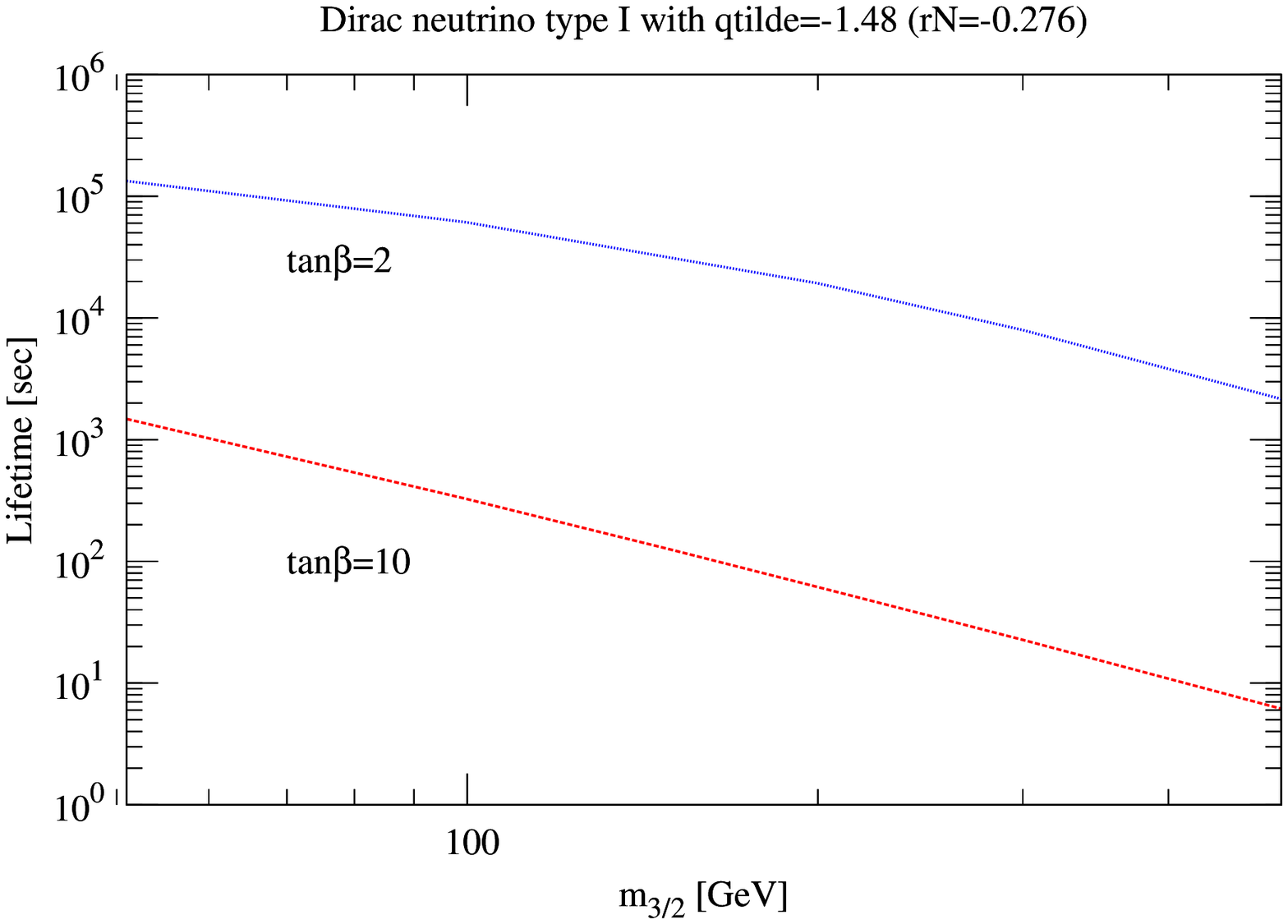}
&
   \includegraphics[width=0.5\textwidth]{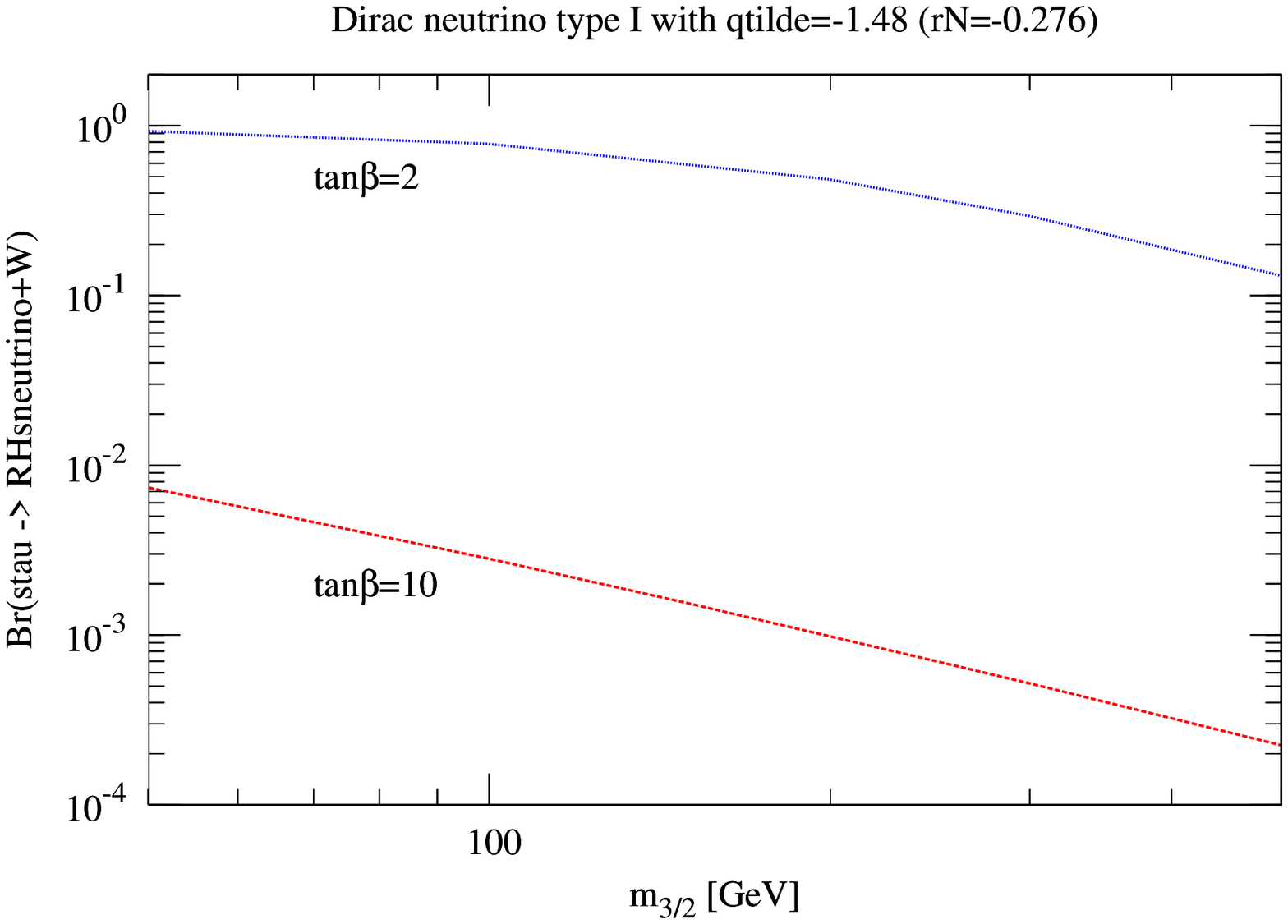}
   \end{tabular}
  \end{center}
 \caption{The lifetime (left) and the branching ratio of stau decay to RH sneutrino
   and W boson (right) for $\tgbeta=2,10$ in the case of Dirac neutrino type I.}
\label{Br_op3}
\end{figure}

In the case of Dirac neutrino, the decay rate of stau to  RH
sneutrino and $W$ gauge boson can be comparable to the one of stau to gravitino and tau lepton,
causing a BBN problem.
The latter decay rate is given in \eq{staulife} and the former one is given by~\cite{aim,bbnmoroi}
 \dis{
 & \Gamma(\tilde{\tau}_1 \to W^- + \tilde{N}_\pm)
  \approx {\sin^2\theta_{\tilde{\tau}}\over 32
    \pi}\bfrac{m^2_{{\tilde\tau}_1}}{m^2_{\tilde{\nu}_L} - m^2_{\tilde{N}_\pm}}^2
          {|\mu\cot\beta-A_\nu^*|^2 m^2_\nu \over
  m_{\tilde{\tau}_1}v^2}\\
&~~~~ \approx 3.3 \times10^{-26}\gev\,\sin^2\theta_{\tilde\tau}
  \bfrac{1\tev}{m_{\tilde{\nu}_L}}^4\bfrac{\mstau}{1\tev}^3\bfrac{|\mu\cot\beta-A_\nu^*|}{1\tev}^2
  \left( m_\nu \over 0.01 {\rm eV} \right)^2 \,
 }
where use is made of $v\simeq 174{\rm GeV}$, $m_\nu$ is the neutrino mass, and  $\theta_{\tilde\tau}$ is the left-right mixing angle of stau, i.e. ${\tilde\tau}_1={\tilde\tau}_R \cos\theta_{\tilde\tau}+{\tilde\tau}_L\sin\theta_{\tilde\tau}$.
Here we used \eq{mnu} and $A_\nu\simeq -2m_{3/2}-0.59M_{1/2}$.
Using this we show the plot of the lifetime of stau and
the branching ratio of stau decay to RH sneutrino and W boson in
Figure~\ref{Br_op3} for $\tgbeta=2,10$ respectively.

As can be seen from Figure 5, for $\tgbeta=10$, one finds that the decay rate
of stau to gravitino and tau lepton (\ref{staulife}) is much larger than the
one for stau to RH sneutrino and $W$ gauge boson. Then, the lifetime of stau is determined by the decay rate to
gravitino and tau lepton so it is less than about 100 sec.
Therefore, in this case, the hadronic particles produced from W boson decay
do not have a BBN problem~\cite{aim,bbnmoroi}. For $\tgbeta$ smaller than 10, the stau
decay rate to RH sneutrino and $W$ boson becomes sizable and thus it would cause the
BBN problem.

On the other hand, the late decay of gravitino to RH sneutrino and neutrino
could cause a BBN problem. But, the BBN constraints on the late decaying gravitino may be avoided if the mass difference
between gravitino and sneutrino is less than about 100 GeV~\cite{Feng:2004zu}.

For the Dirac neutrino Type II with the superpotential
\eq{WdiracII}, the mass range of RH sneutrino is
\dis{
1.49\mgravitino\lesssim  m_{{\tilde N}} < 2.23 \mgravitino. }
 In
this case, RH sneutrino is NLSP and gravitino is LSP as a DM
candidate. The correct relic density of gravitino can be obtained
from the stau decay in some region of heavy stau mass while the
BBN constraint can be avoided in the same way as Dirac neutrino
Type I model discussed above.

In both Dirac neutrino cases, in the region where the non-thermal
production is not enough, the thermal production of gravitino may
give rise to the correct DM of gravitino (or RH sneutrino DM from
gravitino decay) with appropriate reheating temperature around
$10^{9-10}\gev$ (or scaled by $\mgravitino/m_{{\tilde N}}$).

\section{Conclusion}

We have shown that the gauged $U(1)_R$ symmetry naturally realizes
the solution to the $\mu$ problem and accommodates the axion
solution to the strong CP problem. The interplay between the
higher dimensional interaction for singlets and the singlet soft
masses coming from the $U(1)_R$ mediation gives rise to the
stabilization of the singlets at an intermediate scale,
consequently generating the small $\mu$ term from a higher
dimensional interaction.

The gauged $U(1)_R$ symmetry restricts the generated $B\mu$ term
to be larger than the other scalar soft masses of order the
gravitino mass, resulting in $M_{1/2}\gg m_{3/2}$. Thus, we found
that superpartner masses at the EWSB scale in the MSSM sector are
much larger than the mass of gravitino or RH sneutrino. Therefore,
only gravitino or RH sneutrino can be a natural dark matter
candidate.

Depending on whether the Majorana mass term for the RH neutrino
exists,  we considered a different candidate for dark matter:
gravitino for the Majorana neutrino case and RH sneutrino for the
Dirac neutrino case. In both dark matter scenarios, the NLSP in
the MSSM sector is stau. For the stau decaying after the
freezeout, we showed that the correct relic density of dark matter
can be generated by the non-thermal production mechanism through
the stau decay.  At the same time, the BBN constraints on such a
long-lived charged particle can be evaded for the TeV-scale stau
mass. Gravitino and RH sneutrino LSP could be also produced
thermally, by the reheating after inflation and by the decay of
the heavier superparticles in thermal bath, respectively. However,
we have not considered the thermal production because the former
case depends on the reheating temperature and the latter case
depends on the (fine-tuned) enhancement factor of the decay rate.

\section*{Acknowledgments}
K.Y. Choi was partly supported by the Korea Research
Foundation Grant funded by the Korean Government (KRF-2008-341-C00008)
and by the second stage of Brain Korea 21 Project in 2006.
E.J.C. was supported by Korea Neutrino Research
Center through National Research Foundation of Korea Grant
(2009-0083526).

\def\theequation{A.\arabic{equation}}
\setcounter{equation}{0}
\vskip0.8cm
\noindent
{\Large \bf Appendix A: Minimization of the scalar potential for singlets}
\vskip0.4cm
\noindent

In this section, in order to get nonzero singlet VEVs, we focus on the minimization of the scalar potential
for the singlets, $X$ and $Y$.

The F-term potential for singlets $X,Y$ is
\be
V_F=\Big|\frac{2h}{M_P}YH_u H_d+\frac{\kappa}{M_P}X^3\Big|^2+\frac{9\kappa^2}{M^2_P}|Y|^2|X|^4.
\ee

Since the Higgs doublet get small VEVs compared to the singlet VEVs, we ignore the singlet coupling to the Higgs doublets in the F-term potential.
As far as singlet F-terms are negligible and $|X|,|Y|\ll M_P$, the $U(1)_R$ D-term contributes to the singlet scalar potential only through the soft mass terms, which are determined after the moduli stabilization.
Then, adding the singlet soft mass terms to the F-term potential, we obtain the scalar potential for singlets as
\bea
V(X,Y)&=&V_F+m^2_X|X|^2+m^2_Y|Y|^2 \nonumber \\
&\simeq&\frac{\kappa^2}{M^2_P}|X|^6+\frac{9\kappa^2}{M^2_P}|Y|^2|X|^4+m^2_X|X|^2+m^2_Y|Y|^2 +\frac{\kappa}{M_P}A_\kappa Y X^3+{\rm c.c.} \label{spot}
\eea
Writing $X=|X|e^{i\theta_X}$, $Y=|Y|e^{i\theta_Y}$, we find that for $A_\kappa\simeq -2m_{3/2}<0$ and $\kappa>0$, the trilinear terms stabilize one of linear combinations of angles at $3\theta_X+\theta_Y=2n\pi$
with integer $n$. Thus, the other combination of angles becomes a massless axion.
After plugging the minimization condition for the angles into Eq.~(\ref{spot}),
the scalar potential becomes
\bea
V(X,Y)=\frac{\kappa^2}{M^2_P}|X|^6+\frac{9\kappa^2}{M^2_P}|Y|^2|X|^4+m^2_X|X|^2+m^2_Y|Y|^2-\frac{2\kappa |A_\kappa|}{M_P}|Y| |X|^3.
\eea
For the soft mass parameters given in the previous section,
redefining the singlet fields as $x^2=\frac{\kappa|X|^2}{m_{3/2}M_P}$ and $y^2=\frac{\kappa|Y|^2}{m_{3/2}M_P}$,
we rewrite the scalar potential as
\be
V(x,y)=\frac{m^3_{3/2}M_P}{\kappa}\,\Big(x^6+9y^2x^4-\frac{5}{3}x^2+3y^2-4yx^3\Big).
\ee

The extremum conditions for $x$ and $y$ are
\bea
0&=&6x^5+36y^2x^3-\frac{10}{3}x-12yx^2, \\
0&=&18yx^4+6y-4x^3.
\eea
Consequently, we find a minimum at $x\simeq 0.921$ and $y\simeq 0.165$ while $x=y=0$ is a saddle point.
Then, the singlet VEVs are
\bea
|X|\simeq 0.921\sqrt{\frac{m_{3/2}M_P}{\kappa}}, \quad |Y|\simeq 0.165\sqrt{\frac{m_{3/2}M_P}{\kappa}}. \label{singletvevs}
\eea

Expanding the singlets, $X$ and $Y$, around the background VEVs, as $X=\langle X\rangle+\frac{1}{\sqrt{2}}(h_1+i\varphi_2)$
and $Y=\langle Y\rangle +\frac{1}{\sqrt{2}}(h_2+i\varphi_2)$,
we obtain the nonzero mass eigenvalues for singlets:
for real bosons,
\be
M^2_{h_\pm}=\frac{1}{2}\Big(a+b\pm \sqrt{(a-b)^2+4c^2}\Big)
\ee
with
\bea
a&=&\frac{15\kappa^2}{M^2_P}\langle X\rangle^4+\frac{54\kappa^2}{M^2_P}\langle Y\rangle^2 \langle X\rangle^2
-\frac{5}{3}m^2_{3/2}+\frac{6\kappa}{M_P}A_\kappa\langle Y\rangle \langle X\rangle,\\
b&=&\frac{9\kappa^2}{M^2_P}\langle X\rangle^4+3m^2_{3/2}, \\
c&=&\frac{36\kappa^2}{M^2_P}\langle Y\rangle \langle X \rangle^3+\frac{3\kappa}{M_P}A_\kappa \langle X\rangle^2,
\eea
and
\be
M^2_{\varphi_+}=\frac{12\kappa^2}{M^2_P}\langle X\rangle^4+\frac{18\kappa^2}{M^2_P}\langle Y\rangle^2 \langle X\rangle^2
+\frac{4}{3}m^2_{3/2}-\frac{6\kappa}{M_P}A_\kappa \langle Y\rangle \langle X\rangle;
\ee
for Weyl fermions,
\be
M^2_{f_\pm}=\frac{1}{2}\Big(a'+b'\pm \sqrt{(a'-b')^2+4c^{\prime 2}}\Big)
\ee
with
\bea
a'&=&\frac{36\kappa^2}{M^2_P}\langle Y\rangle^2 \langle X\rangle^2+\frac{9\kappa^2}{M^2_P}\langle X\rangle^4,\\
b'&=&\frac{9\kappa^2}{M^2_P}\langle X\rangle^4, \\
c'&=&\frac{18\kappa^2}{M^2_P}\langle Y \rangle \langle X\rangle^3.
\eea
Another combination of the imaginary part is massless and it appears as a Goldstone boson for breaking the PQ symmetry.
For the obtained singlet VEVs (\ref{singletvevs}), we can determine the mass eigenvalues:
for the radial modes, which are almost mass eigenstates due to a small mixing, $M^2_{h_+}=9.67 m^2_{3/2}$ and $M^2_{h_-}=8.39 m^2_{3/2}$; for the massive angular mode, $M^2_{\varphi_+}=12.2m^2_{3/2}$; for Weyl fermions,
$M^2_{f_+}=9.26 m^2_{3/2}$ and $M^2_{f_-}=4.54 m^2_{3/2}$.
Here we note that the radial modes, $h_\pm$ are almost mass eigenstates $h_{2,1}$ due to a small mixing.

\def\theequation{B.\arabic{equation}}
\setcounter{equation}{0}
\vskip0.8cm
\noindent
{\Large \bf Appendix B: Axion for multiple scalar field VEVs}
\vskip0.4cm
\noindent

We identify the axion when multiple scalar fields participate in $PQ$ symmetry breaking.

As shown in Appendix A, a linear combination of angles of singlet scalar fields, $X$ and $Y$, in our model, i.e. $3\theta_X+\theta_Y$, is stabilized by the $A$-term for $YX^3$ term in the superpotential. From the $PQ$-charges of $X$ and $Y$, this combination of angles does not transform under the $U(1)_{PQ}$. So, the orthogonal combination of angles plays a role for the QCD axion.

From the kinetic term for $X,Y$, $-\langle
X\rangle^2(\partial_\mu\theta_X)^2-\langle Y\rangle^2
(\partial_\mu\theta_Y)^2$, we find the canonical axion field as
follows, \be a=\frac{1}{M}\Big(\langle X\rangle a_X-3\langle
Y\rangle a_Y\Big) \ee where $a_X=\frac{\theta_X}{\langle
X\rangle}$, $a_Y=\frac{\theta_Y}{\langle Y\rangle}$ and
$M=\sqrt{9\langle Y\rangle^2+\langle X\rangle^2}$. In the presence
of multiple scalars with VEV $v_i$ and $PQ$-charge $q^i$, the
axion field is generalized \cite{kschoi} to
$a=\frac{1}{M}\sum_{i}a_iq^i v_i$ with
$M=\sqrt{\sum_i(q^iv_i)^2}$.

Then, the axion coupling to the gluon field is given by the following effective Lagrangian,
\be
{\cal L}_{agg}=\frac{a}{f_a}\frac{g^2}{32\pi^2}{\rm tr}(G_{\mu\nu}{\tilde G}^{\mu\nu})
\ee
where $f_a=\frac{M}{{\cal A}}$ with ${\cal A}=\sum_i q^i l_i$ being $U(1)_{PQ}-SU(3)_C-SU(3)_C$ anomaly and $l_i$ being the $SU(3)_C$ quadratic index of a fermion with $PQ$-charge $q^i$. In our case, we obtain the anomaly as ${\cal A}=-3$.
Thus, the axion decay constant is given by $|f_a|=M/|{\cal A}|=\frac{1}{3}\sqrt{9\langle Y\rangle^2+\langle X\rangle^2}$.

\def\theequation{C.\arabic{equation}}
\setcounter{equation}{0}
\vskip0.8cm
\noindent
{\Large \bf Appendix C: PQ symmetry breaking terms and axion solution to strong CP problem}
\vskip0.4cm
\noindent

We must also check other higher dimensional operators which are not $PQ$ symmetric
and thus can potentially spoil the property of the $PQ$ symmetry
for solving the strong CP problem.

If there is a Planck-scale induced non-$PQ$ symmetric term in the potential \cite{pqbreak}:
\be
       V=\frac{1}{M^{2n}_P}\,\phi^{2n+3}(\alpha \phi+\alpha^*\phi^*),
\ee
it gives additional contribution to the axion mass
$m^2 = |\alpha||\phi|^{2n+4}\cos\delta/(M_P^{2n}f_a^2)$ with $\delta$ being the phase of $\alpha$.
In order not to perturb the axion potential term from QCD instanton effect,
this mass must be smaller than about $10^{-5}$ times the usual axion mass ($m_a$):
$m^2 < 10^{-9} m^2_a$.

First, we note that the PQ symmetry is an approximate global symmetry because it is broken explicitly by the Planck-scale
suppressed $U(1)_R$-invariant higher dimensional interactions, e.g. $W=\frac{1}{M^{15}_P}Y^6 X^{12}$.
But, the leading term breaking the $PQ$ symmetry while preserving the $R$ symmetry is given by
$W=\frac{\alpha}{M^{11}_P} (W_0)^4Y^2$. Then, from the F-term potential for $Y$, we obtain the additional term for the axion potential as
\be
\Delta V(a)=\frac{2\kappa\alpha}{M^{12}_P}(W_0)^4 X^{*3}Y+{\rm c.c.}
\ee
Thus, by expanding the above potential around the axion minimum, we get the correction to the axion potential as
\be
\Delta V(a)\simeq m^2_*a^2-\frac{3}{2}f_a m^2_*(\tan\delta)\, a
\ee
where $m^2_*=-\frac{32\kappa|\alpha|}{9f^2_aM^{12}_P}(W_0)^4|X|^3|Y|\cos\delta$ and $e^{-i\delta}=\frac{\alpha}{|\alpha|}$.
Then, from the bound $\langle{\bar\theta}\rangle=\frac{\langle a\rangle}{f_a}< 10^{-9}$,
we obtain
\be
\frac{3}{4}\bigg|\frac{m^2_*\tan\delta}{m^2_a+m^2_*}\bigg|< 10^{-9}
\ee
where $m^2_a=\frac{\Lambda^4_{QCD}}{f^2_a}$. Therefore, for $|m^2_*|\ll m^2_a$, this bound becomes $\frac{3}{4}\frac{|m^2_*|}{m^2_a}\lesssim 10^{-9}$.
For nonzero singlet VEVs given in Eq.~(\ref{singletvevs}), the axion mass correction is $|m^2_*|\sim 0.3\times|\alpha|\Big(\frac{m_{3/2}}{M_P}\Big)^3m^2_{3/2}\sim 10^{-26}{\rm eV}^2$ for $m_{3/2}\sim 100{\rm GeV}$.
Compared to the axion mass bound, $m_a=(0.6\times 10^7{\rm GeV}/f_a){\rm eV}\gtrsim 0.6\times 10^{-5}{\rm eV}$ for $f_a<10^{12}{\rm GeV}$, the axion mass correction is negligible.

\def\theequation{D.\arabic{equation}}
\setcounter{equation}{0}
\vskip0.8cm
\noindent
{\Large \bf Appendix D: R-parity violating terms}
\vskip0.4cm
\noindent

In this appendix, we consider the R-parity violating terms induced after the $R$-symmetry breakdown.
We focus on the Majorana neutrino case with $n=2$.

The effective R-parity violating terms are generated by the following higher dimensional interactions:
\be
\frac{1}{M^6_P} YX^5 L Q D, \ \  \frac{1}{M^6_P} YX^5 L L E,
\ \ \frac{1}{M^6_P}(W_0)^2 UDD, \ \ \frac{1}{M^7_P}W_0Y^2X^2 LH_u.\nonumber
\ee
However, after the superpotential and the singlets develop a nonzero VEV,
the induced R-parity violating couplings are negligible. Therefore, it is possible to have a stable LSP.

Moreover, we also note that the following terms are allowed:
\be
\frac{1}{M^9_P}(W_0)^2X^2QQQL, \quad \frac{1}{M^{12}_P}(W_0)^3YX UUDE.\nonumber
\ee
Thus, the B/L violating dim-5 operators are negligible so the proton stability is also justified.

\end{document}